\documentclass[%
 reprint,
superscriptaddress,
 amsmath,
 amssymb,
 aps,
prc,
]{revtex4-2}

\usepackage{hyperref}
\usepackage{amsmath}
\usepackage{amssymb}
\usepackage{graphicx}
\usepackage{mathrsfs}
\usepackage{color}
\usepackage{bm}
\usepackage{ulem}

\DeclareRobustCommand{\erase}{\bgroup\markoverwith{\textcolor{red}{\rule[.5ex]{2pt}{1.pt}}}\ULon}

\begin{document}

\title{Structure of multi-$\Lambda$ hypernuclei with a Skyrme-type $\Lambda\Lambda$ interaction constrained by data on double-$\Lambda$ hypernuclei and neutron stars}

\author{Yusuke Tanimura}
\affiliation{Department of Physics and Origin of Matter and Evolution of Galaxies (OMEG) Institute, Soongsil University, Seoul 06978, Korea}

\author{Chang Ho Hyun}
\affiliation{Department of Physics Education, Daegu University, Gyeongsan 38453, Korea}

\author{Myung-Ki Cheoun}
\affiliation{Department of Physics and Origin of Matter and Evolution of Galaxies (OMEG) Institute, Soongsil University, Seoul 06978, Korea}

\date{\today}

\begin{abstract}
We investigate multi-$\Lambda$ hypernuclear systems with Skyrme-type $\Lambda\Lambda$
interactions constrained by the data on double-$\Lambda$ hypernuclei and neutron stars.
The roles of the repulsive $p$-wave and density-dependent terms in the $\Lambda\Lambda$ interaction
are examined by considering the homogeneous hyperonic matter around the normal 
density and finite multi-$\Lambda$ hypernuclei
within the spherical Hartree-Fock approach.
In homogeneous matter, the $\Lambda$ chemical potential and
corresponding $\Lambda$ drip point depend strongly on the repulsive $p$-wave term,
while the effect of density-dependent term is relatively weak in the density range
relevant to finite nuclei.
In the multi-$\Lambda$ hypernuclei built on doubly closed stable cores from light to heavy
systems, $\Lambda$ radius, separation energy and single-particle structure 
show a clear dependence on the repulsive $p$-wave interaction, and this dependence becomes
stronger as the number of $\Lambda$ hyperons increases.
A second and distinct effect appears near the $\Lambda$ drip line: when the last occupied
$\Lambda$ orbit approaches the continuum, the repulsive $p$-wave term shifts the 
state upward and can produce a weakly bound state with an extended radial distribution.
As a result, $\Lambda$ radius can increase rapidly near the threshold. 
This threshold effect should be distinguished from the moderate enhancement of the dependence 
on $p$-wave interaction with increasing number of $\Lambda$ hyperons.
These results indicate that the multi-$\Lambda$ hypernuclei are particularly useful
for isolating the role of $p$-wave $\Lambda\Lambda$ interacion around the normal density,
whereas the density-dependent term is expected to be more important interaction in the
high-density domain relevant to neutron stars.
\end{abstract}


\maketitle


\section{Introduction}

The $\Lambda\Lambda$ interaction plays an essential role in the structure of 
double-$\Lambda$ and multi-$\Lambda$ hypernuclei, and is also relevant to the
equation of state of strange hadronic matter. 
Experimental information on the $S=-2$ sector is still limited to a small number of 
double-$\Lambda$ hypernuclear events, although substantial progress has been made in hyper 
nuclear spectroscopy and in the study of double-strangeness systems \cite{HaTa06,HiNa18}. 
Theoretical studies of multi-$\Lambda$ systems have a long history and have provided 
useful insight into the shell structure, density distributions, and possible drip-line behavior of 
strangeness-rich finite nuclei~\cite{BIM,MIB83,IBM85,Rufa90,MZ93,Sch94,LaYa97,La98,Cugnon00,Hiyama02, HKYMR10,MiHa12,MCH09,YaMoRi10,MiCh11,Gal11,Schulze14,KMGR15,MKG17,GBKM18,Ta19,Guo22,Su24,LCZR24,ZhHiSa25,TaHyCh26}. 
In particular, mean-field approaches make it possible to investigate
multi-$\Lambda$ hypernuclei systematically over a wide mass region, including
heavy systems that are difficult to treat within few-body aproaches.

Among the mean-field approaches, both relativistic mean-field (RMF) models
and Skyrme-type energy density functionals have been widely used. 
In RMF models, the effective interaction is described in terms of scalar-
and vector-meson exchange, supplemented by nonlinear meson self-interactions
or density-dependent couplings \cite{Rufa90,MZ93,Sch94,Ta19}. 
In Skyrme-type models, by contrast, the interaction is represented by an
effective zero-range interaction containing momentum-dependent $s$- and
$p$-wave terms together with density-dependent contributions  \cite{LaYa97,La98,Cugnon00,MiHa12,MCH09,Schulze14,KMGR15,MKG17,GBKM18,Guo22,Su24,LCZR24,ZhHiSa25,TaHyCh26}. 
The covariant RMF framework is particularly advantageous in extrapolations to
high density, where causality is naturally preserved, whereas the zero-range
Skyrme form leads to a simple and flexible energy density functional (EDF) that is
well suited to systematic finite-nucleus calculations.

A major difficulty in studies of multi-$\Lambda$ hypernuclei is that the
$\Lambda\Lambda$ interaction remains only weakly constrained. 
The currently available double-$\Lambda$ hypernuclear data mainly constrain
the low-energy $s$-wave sector, while the $p$-wave part is essentially unknown. 
Recent Skyrme-Hartree-Fock-Bogoliubov (HFB) calculations by Zhang \textit{et al.} \cite{ZhHiSa25}
have illustrated this
point clearly: by varying the strength of the $p$-wave $\Lambda\Lambda$
interaction, they found sizable effects on the shell structure, drip-line behavior, 
and density distribution of multi-$\Lambda$ hypernuclei, especially for neutron-rich O and Ca cores. 
Their results show that the predicted properties of
multi-$\Lambda$ systems can depend strongly on how the poorly known $\Lambda\Lambda$ sector is chosen.

In our previous work \cite{TaHyCh26}, we constructed a Skyrme-type $\Lambda\Lambda$
interaction including the standard $s$-wave and $p$-wave terms together with
a density-dependent term that simulates an effective $N\Lambda\Lambda$ three-body force. 
The $s$-wave parameters were constrained by data on double-$\Lambda$
hypernuclei supplemented by pseudodata from core+$2\Lambda$ three-body
calculations, including somewhat heavier systems for which no experimental
data are presently available. 
We showed that such additional pseudodata are important for determining the
two $s$-wave parameters simultaneously in a stable manner. 
We also examined the roles of the $p$-wave and density-dependent terms in
neutron-star matter, and found that their repulsive contributions can bring
the equation of state into better agreement with current mass-radius observations. 
This framework therefore provides a more tightly constrained $\Lambda\Lambda$
functional than in earlier exploratory studies of multi-$\Lambda$ hypernuclei.

With this constrained interaction at hand, it is natural to ask how the
$p$-wave and density-dependent terms manifest themselves in finite multi-$\Lambda$ hypernuclei. 
The present work addresses this question in the density regime around normal
nuclear density, where finite multi-$\Lambda$ systems provide a complementary
testing ground to the high-density environment of neutron stars. 
Our main interest is the role of the $p$-wave $\Lambda\Lambda$ term in systems
where $\Lambda$ hyperons occupy not only the $1s$ orbit but also the $1p$ and higher shells. 
In contrast to the recent HFB study that emphasized 
neutron-rich O and Ca cores and the associated pairing and continuum effects \cite{ZhHiSa25}, 
we focus primarily on
multi-$\Lambda$ hypernuclei built on doubly closed stable cores spanning a wide
mass range, from light to heavy nuclei. 
Within this setting, the effects of the $\Lambda\Lambda$ interaction on
$\Lambda$ radii, separation energies, shell structure, and drip-line behavior
can be examined in a transparent manner.

Another motivation of the present study is that finite multi-$\Lambda$
hypernuclei probe a density region close to the normal saturation density,
where the nucleonic core is expected to remain relatively stable, while the
$\Lambda$ sector becomes increasingly sensitive to the $\Lambda\Lambda$
interaction as more $\Lambda$ hyperons are added. 
This makes such systems especially suitable for isolating the role of the
$p$-wave term, which does not contribute to the ground states of ordinary
double-$\Lambda$ hypernuclei but can become important once $\Lambda$ hyperons
occupy the $p$ shell and higher orbitals. 
We therefore investigate both the bulk properties of hyperonic matter around
normal density and the finite-nucleus manifestations of the constrained
$\Lambda\Lambda$ functional in spherical Hartree-Fock (HF) calculations. 
Exploratory results for neutron-rich systems are presented separately in the Appendix.

The remainder of this paper is organized as follows. 
Section~\ref{sec:model} outlines the Skyrme-type EDF used in this work, 
with emphasis on the structure of the $\Lambda\Lambda$ interaction. 
In Sec.~\ref{sec:results}, we first examine hyperonic matter around the normal density 
to clarify the roles of the $p$-wave and density-dependent terms, 
and then turn to spherical HF calculations of finite multi-$\Lambda$ hypernuclei 
with doubly-closed stable cores, 
for which the effects of the $\Lambda\Lambda$ interaction on 
$\Lambda$ radii, separation energies, shell structure, and drip-line behavior 
are examined. 
Section~\ref{sec:summary} summarizes the main results and gives an outlook, 
while exploratory calculations for neutron-rich systems are presented in Appendix~\ref{app:n-rich}.

\section{Model: Skyrme energy density functional}
\label{sec:model}

The present calculations are based on the Korea-IBS-Daegu-SKKU (KIDS) Skyrme-type
EDF for multi-$\Lambda$ hypernuclear systems.
The total energy is written as
\begin{align}
E = \int d^3r\, \mathcal{E},
\end{align}
where the total energy density is decomposed into
\begin{align}
\mathcal{E} = \mathcal{E}_{NN} + \mathcal{E}_{N\Lambda} + \mathcal{E}_{\Lambda\Lambda}.
\end{align}
Here, $\mathcal{E}_{NN}$ denotes the nucleonic energy density,
$\mathcal{E}_{N\Lambda}$ includes the kinetic-energy density of $\Lambda$ hyperons
and the effective $N\Lambda$ interaction,
and $\mathcal{E}_{\Lambda\Lambda}$ represents the $\Lambda\Lambda$ contribution.

For the $\Lambda\Lambda$ sector, we employ the Skyrme-type EDF
\cite{La98}
\begin{align}
\mathcal{E}_{\Lambda\Lambda}
&=
\frac{1}{4}\lambda_0 \rho_\Lambda^2
+\frac{1}{8}(\lambda_1+3\lambda_2)\rho_\Lambda\tau_\Lambda
\nonumber
\\
&\quad
+\frac{3}{32}(\lambda_2-\lambda_1)\rho_\Lambda \bm\nabla^2 \rho_\Lambda
+\frac{1}{4}\lambda_3 \rho_\Lambda^2 \rho_N ,
\end{align}
where $\rho_\Lambda$ and $\tau_\Lambda$ are the number density
and kinetic-energy density of $\Lambda$ hyperons, respectively,
and $\rho_N$ is the total nucleon density.
The terms proportional to $\lambda_0$ and $\lambda_1$ represent the
$s$-wave part of the $\Lambda\Lambda$ interaction,
the $\lambda_2$ term corresponds to the $p$-wave part,
and the density-dependent $\lambda_3$ term simulates
an effective repulsive $N\Lambda\Lambda$ three-body force 
and any other many-body correlations.

The $NN$ and $N\Lambda$ sectors are taken from the KIDS and corresponding
Y4 parameter sets \cite{GHHY22,CHHC22}, respectively.
All HF calculations in the present work are performed
under spherical symmetry with the equal-filling approximation.
The center-of-mass correction is taken into account perturbatively
after variation by subtracting the expectation value of the
center-of-mass kinetic energy from the total energy.

As in Ref.~\cite{TaHyCh26}, the parameter sets used below are denoted as
KIDS-(A/D)-Y4-LL4$(\lambda_2,\lambda_3)$, or simply
LL4$(\lambda_2,\lambda_3)$ when no confusion arises.
Here KIDS-A or KIDS-D specifies the $NN$ functional, Y4 denotes the
corresponding $N\Lambda$ parameter set, and LL4 indicates that all four
parameters in the $\Lambda\Lambda$ sector are included.
The values of $\lambda_2$ and $\lambda_3$ are given in units of
MeV\,fm$^5$ and MeV\,fm$^6$, respectively.
For each fixed value of $\lambda_3$, the $s$-wave parameters
$\lambda_0$ and $\lambda_1$ have been refitted to double-$\Lambda$
hypernuclear data, while $\lambda_2$ is varied independently because it
does not contribute to the ground states of ordinary double-$\Lambda$
hypernuclei.
Thus LL4$(0,\lambda_3)$ serves as the reference set without the
$p$-wave $\Lambda\Lambda$ term.

The ranges of $\lambda_2$ and $\lambda_3$ considered in the present work 
should be regarded as a physically motivated parameter scan 
rather than statistically rigorous confidence intervals~\cite{TaHyCh26}.
For the density-dependent term, we restrict ourselves to non-negative values 
of $\lambda_3$, since this term is introduced as an effective repulsive 
$N\Lambda\Lambda$ contribution. 
The upper end of the $\lambda_3$ range is then chosen on the basis of 
the quality of this refit to the double-$\Lambda$ data; 
this practical criterion inevitably contains some arbitrariness.

The parameter $\lambda_2$, on the other hand, is not constrained by the 
ground-state properties of double-$\Lambda$ hypernuclei, 
because the $p$-wave contribution vanishes for the $(1s)^2$ configuration. 
Its lower acceptable range is therefore guided by neutron-star constraints: 
too small a repulsive $\lambda_2$ leads to an insufficiently stiff 
hyperonic equation of state and fails to support massive neutron stars. 
At the same time, the neutron-star mass-radius curves were found to show 
a saturation tendency as $\lambda_2$ increases, so that the present 
neutron-star observables do not provide a sharp upper bound on $\lambda_2$. 
We therefore adopt $\lambda_2=600~{\rm MeV\,fm}^5$ as a practical upper end 
of the parameter scan.

Consequently, statements below about a ``strong'' or ``weak'' dependence 
on $\lambda_2$ or $\lambda_3$ should be interpreted relative to the 
parameter ranges explored in this work.

\section{Results}
\label{sec:results}

Before discussing the individual finite systems, 
it is useful to distinguish two different types of $\lambda_2$ dependence. 
First, the sensitivity of observables such as $r_\Lambda$ and $S_\Lambda$ 
generally increases with $N_\Lambda$, 
because the $\Lambda\Lambda$ contribution accumulates as more $\Lambda$ hyperons are added. 
Second, only near the $\Lambda$ drip line can a rapid change of $r_\Lambda$ occur: 
when the last occupied $\Lambda$ orbit approaches the continuum, 
the repulsive $p$-wave term shifts it upward and produces a weakly bound state 
with an extended radial distribution. 
The former is a generic many-$\Lambda$ effect, 
whereas the latter is a threshold effect associated with the last occupied orbit.

\subsection{Double-$\Lambda$ hypernuclei}

\begin{figure}
\includegraphics[width=\linewidth]{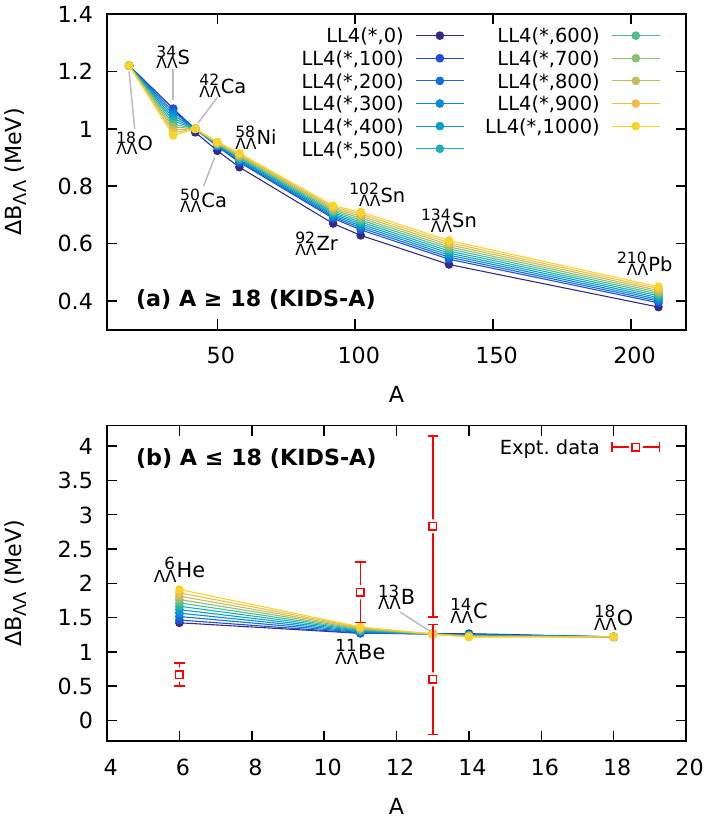}
\caption{Calculated $\Delta B_{\Lambda\Lambda}$ values 
for mass numbers (a) $A\geq 18$ and (b) $A\leq 18$, 
obtained using KIDS-A-Y4-LL4($*,\lambda_3$) parameter sets 
with $\lambda_3$ ranging from 0 to 1000 MeV\,fm$^6$. 
Experimental data for $^{6}_{\Lambda\Lambda}$He and $^{11}_{\Lambda\Lambda}$Be are 
taken from Refs.~\cite{Ahn13,Takahashi01,Ek19}. 
The data for $^{13}_{\Lambda\Lambda}$B, 
$\Delta B_{\Lambda\Lambda} = 0.6 \pm 0.8$~MeV and 
$2.83 \pm 1.18({\rm stat.}) \pm 0.14({\rm syst.})$~MeV, 
are taken from Ref. \cite{Aoki09} and \cite{He25}, respectively. }
\label{fig:DelB_LL-A-2}
\end{figure}

Figure~\ref{fig:DelB_LL-A-2} shows the calculated values of the 
two-$\Lambda$ correlation energy $\Delta B_{\Lambda\Lambda}$ 
for double-$\Lambda$ hypernuclei with 
(a) $A\geq 18$ and (b) $A\leq 18$. 
The results are obtained with the KIDS-A-Y4-LL4($*,\lambda_3$) 
parameter sets, where $\lambda_3$ is varied from 
0 to 1000 MeV\,fm$^6$. 
The two-$\Lambda$ correlation energy is defined as
\begin{align}
\Delta B_{\Lambda\Lambda}\left( ^A_{\Lambda\Lambda} Z \right) &= 
B_{\Lambda\Lambda}\left( ^A_{\Lambda\Lambda} Z \right) 
- 2B_{\Lambda}\left( ^{A-1}_{\Lambda} Z \right), 
\end{align}
where
\begin{align}
B_\Lambda\left(^A_\Lambda Z\right) 
&= B\left(^A_\Lambda Z\right) - B\left(^{A-1}Z\right),
\\
B_{\Lambda\Lambda}\left(^A_{\Lambda\Lambda} Z\right) 
&= B\left(^A_{\Lambda\Lambda} Z\right)
 - B\left(^{A-2} Z\right)
\end{align}
are the single- and double-$\Lambda$ binding energies, respectively.

Since the $p$-wave $\lambda_2$ term does not contribute to the 
ground-state binding energies of ordinary double-$\Lambda$ hypernuclei, 
in which the two $\Lambda$ hyperons occupy the $1s$ orbit, 
the value of $\lambda_2$ is not specified in this figure. 
For each value of $\lambda_3$, the $s$-wave parameters 
$\lambda_0$ and $\lambda_1$ are taken from the refit to the double-$\Lambda$ hypernuclear dataset. 
The fitted values of $\lambda_0$ and $\lambda_1$ for the individual 
parameter sets are listed in the Supplementary Data of Ref.~\cite{TaHyCh26}.

The smooth mass dependence of $\Delta B_{\Lambda\Lambda}$ 
indicates that, for each fixed value of $\lambda_3$, 
the refitted $s$-wave parameters $\lambda_0$ and $\lambda_1$ 
provide a well-behaved description of ordinary double-$\Lambda$ ground states. 
Thus, the part of the $\Lambda\Lambda$ functional that is active 
in the $(1s)^2$ configuration is already constrained at the level 
needed for the present finite-nucleus study. 
Since the $p$-wave parameter $\lambda_2$ is inactive in such 
double-$\Lambda$ ground states, it cannot be constrained by these data. 
The $\lambda_2$ dependence found below in multi-$\Lambda$ systems 
should therefore be interpreted as a genuine many-$\Lambda$ effect 
associated with the occupation of the $1p$ and higher shells, 
rather than as a residual ambiguity of the double-$\Lambda$ fit.

\subsection{Homogeneous hyperonic matter}

Before turning to finite multi-$\Lambda$ hypernuclei, 
it is useful to examine the properties of homogeneous matter around the normal nuclear density. 
The energy per baryon of infinite matter is given as 
\begin{align}
  \frac{E}{A}(\rho_n,\rho_p,\rho_\Lambda) &= \frac{{\cal E}(\rho_n,\rho_p,\rho_\Lambda)}{\rho_B}, 
\end{align}
where $\rho_B = \rho_n+\rho_p+\rho_\Lambda$ is the total baryon density.
The chemical potential of each baryon species ($i=n,\ p$, or $\Lambda$) is given by 
\begin{align}
\mu_i &= \frac{\partial{\cal E}}{\partial\rho_i}. 
\end{align}
To mimic the environment of multi-$\Lambda$ hypernuclei, 
we consider isospin-symmetric hyperonic matter near the normal density, 
$\rho_n=\rho_p=\rho_N/2$ with $\rho_N=\rho_0=0.16$~fm$^{-3}$.

\begin{figure}
\includegraphics[width=\linewidth]{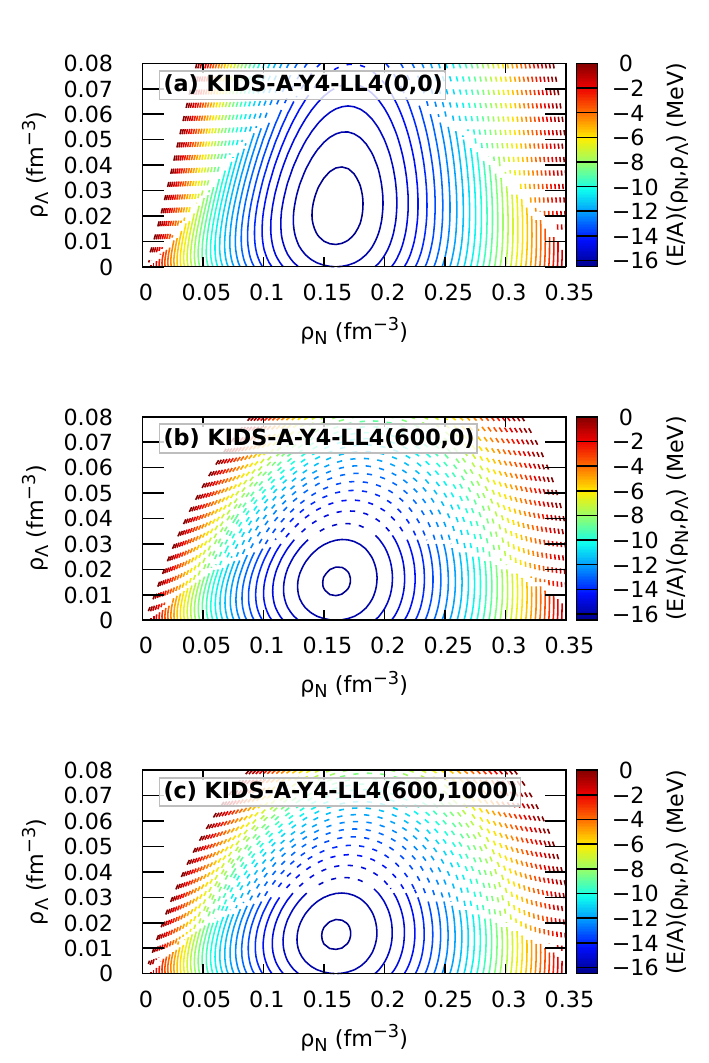}
\caption{Contour plots of the energy per baryon $E/A$ in the $(\rho_N,\rho_\Lambda)$ plane, where $\rho_N=\rho_n+\rho_p$ and $\rho_n=\rho_p=\rho_N/2$, 
for (a) KIDS-A-Y4-LL4(0,0), (b) KIDS-A-Y4-LL4(600,0), and (c) KIDS-A-Y4-LL4(600,1000).
The contours are drawn from $-16.5$~MeV to 0 MeV every $0.5$ MeV. 
The contours are drawn by solid and dashed lines 
in the region where $\mu_\Lambda\leq 0$ and $> 0$, respectively.}
\label{fig:EbyA-Matter}
\end{figure}

Figure~\ref{fig:EbyA-Matter} shows contour plots of the energy per baryon $E/A$ 
in the $(\rho_N,\rho_\Lambda)$ plane for (a) KIDS-A-Y4-LL4(0,0), (b) KIDS-A-Y4-LL4(600,0), and (c) KIDS-A-Y4-LL4(600,1000) parameter sets. 
The contours are drawn by solid and dashed lines 
in the region where $\mu_\Lambda\leq 0$ and $> 0$, respectively.

A minimum of $E/A$ is found at $(\rho_N,\rho_\Lambda)\approx(\rho_0,0.02~\mathrm{fm}^{-3})$ for both values of $\lambda_2$. 
In the low-$\rho_\Lambda$ region, the $N\Lambda$ interaction dominates
 over the $\Lambda\Lambda$ interaction, and therefore the $\lambda_2$ dependence of $E/A$ is weak. 
As $\rho_\Lambda$ increases, however, the $\Lambda\Lambda$ contribution becomes more important, 
and the repulsive effect of the $\lambda_2$ term becomes clearly visible.

A comparison between Figs. \ref{fig:EbyA-Matter}(b) and \ref{fig:EbyA-Matter}(c) shows
that the $\lambda_3$ dependence of $E/A$ is relatively small. 
This is mainly because the $s$-wave parameters $\lambda_0$ and $\lambda_1$ 
are refitted to the double-$\Lambda$ hypernuclear data for each value of $\lambda_3$, 
so that the matter properties are consistent with the dataset
and remain insensitive to $\lambda_3$ at least around the normal density.

For a fixed value of $\rho_\Lambda$, the value of $\rho_N$ that minimizes $E/A$ 
remains close to the saturation density $\rho_0$ in the region where $\mu_\Lambda<0$. 
This indicates that the normal nuclear core of a multi-$\Lambda$ system 
is not expected to undergo a substantial compression or expansion 
by the addition of $\Lambda$ hyperons.

\begin{figure}
\includegraphics[width=\linewidth]{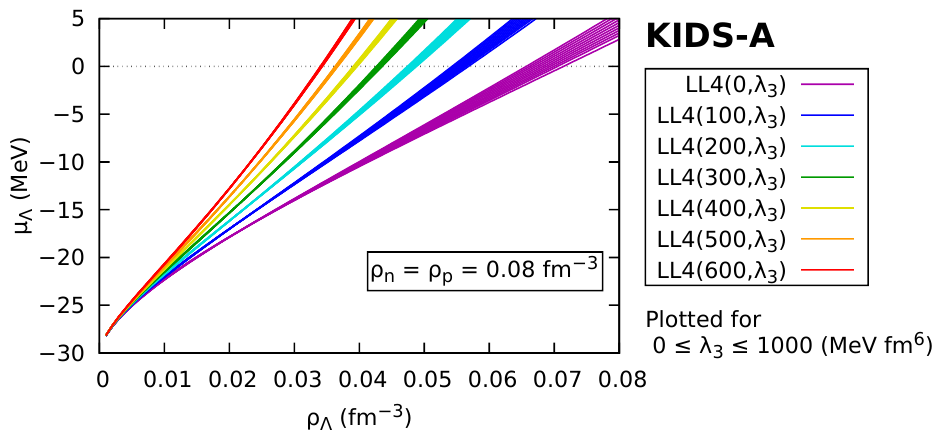}
\caption{$\Lambda$ hyperon Fermi energy $\mu_\Lambda$ as a function of $\rho_\Lambda$ with $\rho_n=\rho_p=0.08$~fm$^{-3}$ for KIDS-A-Y4-LL4$(\lambda_2,\lambda_3)$ parameter sets 
with $\lambda_3=0, 100, 200, \dots, 1000$ MeV\,fm$^{6}$ \cite{TaHyCh26}. }
\label{fig:muL-Matter}
\end{figure}

In Fig.~\ref{fig:muL-Matter} we plot 
the $\Lambda$ Fermi energy $\mu_\Lambda$ as a function of 
$\rho_\Lambda$ at fixed nucleon densities $\rho_n=\rho_p=0.08~{\rm fm}^{-3}$
for KIDS-A-Y4-LL4($\lambda_2$, $\lambda_3$) parameter sets.
The $\lambda_3$ dependence of $\mu_\Lambda$ within the range considered here is relatively weak, 
and both $\lambda_2$ and $\lambda_3$ dependences become weaker as $\rho_\Lambda$ approaches zero.
In the limit $\rho_\Lambda\to 0$, $\mu_\Lambda$ reduces to the mean-field
potential of a single $\Lambda$ in nuclear matter, corresponding to the
empirical depth $U_\Lambda(\rho_0)\simeq -30$~MeV.
As expected, all parameter sets approach this value, because the
$N\Lambda$ sector has been adjusted to single-$\Lambda$ hypernuclear data.
With a stronger $\Lambda\Lambda$ repulsion, $\mu_\Lambda$ becomes larger at a given $\rho_\Lambda$.

We refer to the condition $\mu_\Lambda=0$ as the $\Lambda$ drip point in homogeneous matter, 
in analogy with the drip-line condition in finite nuclei. 
At $\rho_n = \rho_p = 0.08$ fm$^{-3}$, this point is located at
$\rho_\Lambda \approx 0.07$ fm$^{-3}$ for $\lambda_2 = 0$ and 
at $\rho_\Lambda \approx 0.03$ fm$^{-3}$ for $\lambda_2 = 600$ MeV\,fm$^5$.

The present analysis of homogeneous hyperonic matter 
therefore clarifies characteristic features of nucleon-$\Lambda$ matter 
around the normal density. 
The nucleonic density tends to remain close to the saturation density, 
whereas the $\Lambda$ chemical potential and the corresponding 
$\Lambda$ drip point are strongly affected by $\lambda_2$. 
Within the parameter range explored in this work, 
the variation induced by $\lambda_3$ is smaller than that induced by $\lambda_2$ 
in the density region relevant to finite multi-$\Lambda$ hypernuclei. This bulk-matter behavior provides a useful basis for interpreting 
the finite multi-$\Lambda$ hypernuclei discussed below.

\subsection{Multi-$\Lambda$ systems}

We now turn to finite multi-$\Lambda$ hypernuclei calculated within the
spherical HF framework.
A system is regarded as bound when all added $\Lambda$ hyperons can be
accommodated in bound HF single-particle states; if the required filling
exceeds the bound spectrum, the system is regarded as beyond the
$\Lambda$ drip line and is not included in the plots.
Continuum states are therefore not used in the present HF calculations.
The discussion below first uses the $^{16}$O core to illustrate the
mechanism of the $p$-wave effect, and then extends the same viewpoint 
to heavier closed-core systems.

\begin{figure*}[t]
\includegraphics[width=.9\linewidth]{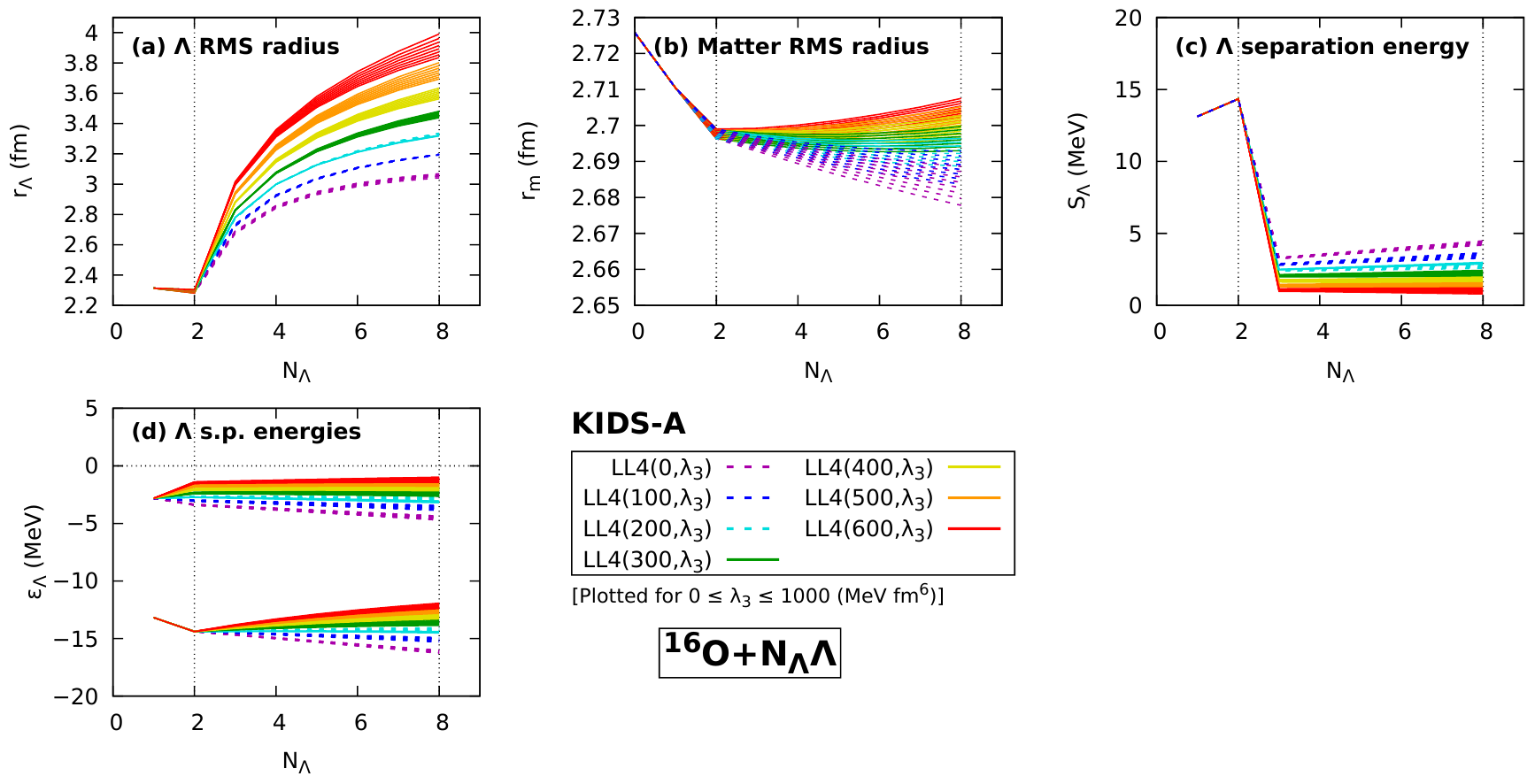}
\caption{
HF results for $^{16}$O-core multi-$\Lambda$ hypernuclei, 
calculated with the KIDS-A-Y4-LL4$(\lambda_2,\lambda_3)$
parameter sets.
Panels show (a) the $\Lambda$ root-mean-square (rms) radius $r_\Lambda$, (b) the matter rms
radius $r_m$, (c) the $\Lambda$ separation energy $S_\Lambda$, and (d) the
$\Lambda$ single-particle energies as functions of the number of
$\Lambda$ hyperons $N_\Lambda$.
The line color distinguishes the value of $\lambda_2$, while the spread of
curves with the same color reflects the variation of $\lambda_3$ over the
range shown in the plot.
Dashed curves correspond to parameter sets disfavored by observations of
massive ($\sim2M_\odot$) neutron stars \cite{TaHyCh26}.
The vertical dotted lines indicate the closures of the $1s$ and $1p$
$\Lambda$ shells at $N_\Lambda=2$ and 8, respectively.
}
\label{fig:O16+mul-L}
\end{figure*}

Figure~\ref{fig:O16+mul-L} shows the HF results for multi-$\Lambda$ hypernuclei 
with $^{16}$O core obtained using KIDS-A-Y4-LL4($\lambda_2,\lambda_3$) sets. 
The panels show 
(a) the $\Lambda$ root-mean-square (rms) radius $r_\Lambda$, 
(b) the matter rms radius $r_m$, 
(c) the $\Lambda$ separation energy $S_\Lambda$, and 
(d) the $\Lambda$ single-particle energies 
as functions of the number of $\Lambda$ hyperons $N_\Lambda$.
The line color distinguishes the value of $\lambda_2$, while the spread of
curves with the same color reflects the variation of $\lambda_3$ over the
range shown in the figure.
Dashed curves correspond to parameter sets disfavored by observations of
massive ($\sim2M_\odot$) neutron stars \cite{TaHyCh26}.
Figure~\ref{fig:O16+mul-L} demonstrates in a finite system the same features as homogeneous matter. 
The matter radius changes only weakly with $N_\Lambda$, 
showing that the core-polarization effect remains small, 
whereas $r_\Lambda$, $S_\Lambda$, and the $1s$-$1p$ shell spacing 
display an increasingly visible $\lambda_2$ dependence as $N_\Lambda$ increases. 
This gradual trend reflects the growing contribution of the $\Lambda\Lambda$ interaction 
as more $\Lambda$ hyperons are accommodated in the system. 
In addition, a distinct threshold effect appears near the $\Lambda$ drip line: 
once the last occupied $1p$ orbit approaches the continuum, 
the repulsive $\lambda_2$ term pushes it upward and $r_\Lambda$ increases rapidly. 
This rapid near-threshold variation should be distinguished 
from the more gradual enhancement of the $\lambda_2$ dependence with increasing $N_\Lambda$.

As seen in Fig.~\ref{fig:O16+mul-L}(b), 
the matter radius first decreases when a few $\Lambda$ hyperons 
are added to the normal core. 
This is the usual gluelike effect of the attractive $N\Lambda$ interaction. 
When more $\Lambda$ hyperons are added, however, 
the $\Lambda$ radius shown in Fig.~\ref{fig:O16+mul-L}(a) 
becomes substantially larger than $r_m$. 
The spatially extended $\Lambda$ distribution then produces an attractive 
mean field for the nucleons over a larger radial region, 
and the matter radius starts to increase slightly.

The $\Lambda$ separation energy shown in 
Fig.~\ref{fig:O16+mul-L}(c) is independent of the 
$p$-wave coupling $\lambda_2$ as long as only the $1s$ $\Lambda$ orbit is occupied. 
A $\lambda_2$ dependence appears only after the $\Lambda$ hyperons 
start to occupy the $1p$ orbit. 
In this case, the $\lambda_2$ term contributes through
relative $p$-wave components involving the occupied $1p$ orbit. 
Correspondingly, the single-particle energies in 
Fig.~\ref{fig:O16+mul-L}(d) show that the $1s$-$1p$ spacing 
becomes smaller as the number of $\Lambda$ hyperons in the $1p$ orbit 
increases for larger values of $\lambda_2$.

\begin{figure*}
\includegraphics[width=.9\linewidth, page=1]{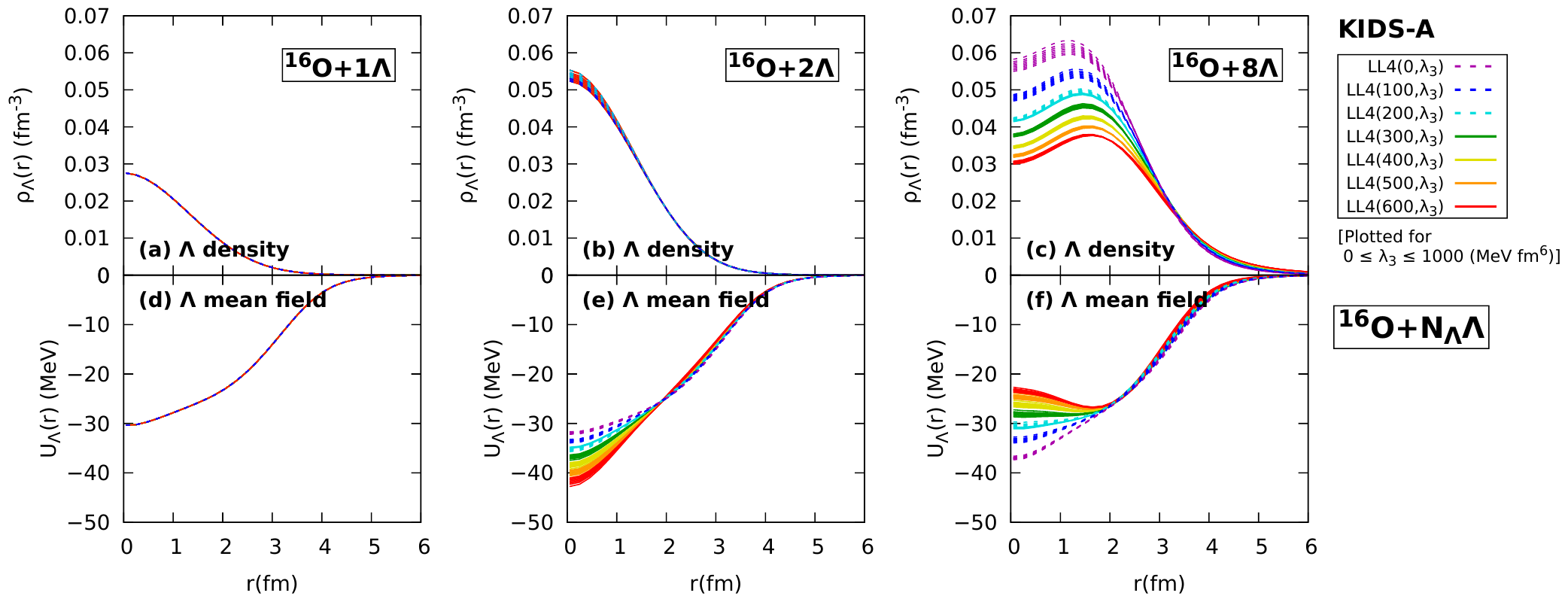}
\caption{
$\Lambda$ density distributions $\rho_\Lambda(r)$ 
and the local central part of the $\Lambda$ mean-field potential 
$U_\Lambda(r)$ in $^{16}{\rm O}+N_\Lambda\Lambda$ 
calculated with the KIDS-A-Y4-LL4$(\lambda_2,\lambda_3)$ 
parameter sets. 
The results are shown for 
(a,d) $N_\Lambda=1$, 
(b,e) $N_\Lambda=2$, and 
(c,f) $N_\Lambda=8$. 
The line color distinguishes the value of $\lambda_2$, 
while, for each $\lambda_2$, results with 
$0\le \lambda_3\le 1000~{\rm MeV\,fm}^6$ are overlaid. 
The dashed curves correspond to parameter sets disfavored by observations of massive ($\sim 2M_\odot$) neutron stars \cite{TaHyCh26}.}
\label{fig:densities-O16+mul-L}
\end{figure*}

Figure~\ref{fig:densities-O16+mul-L} shows the $\Lambda$ density distributions 
and the local central part of the $\Lambda$ mean-field potential 
for $^{16}{\rm O}+N_\Lambda\Lambda$. 
For the single-$\Lambda$ system, all curves coincide, as expected, 
because the $\Lambda\Lambda$ interaction is absent. 
The density distribution and the mean field are therefore determined only by 
the $N\Lambda$ sector.

For $N_\Lambda=2$, the two $\Lambda$ hyperons occupy the $1s$ orbit. 
The density distribution is almost independent of $\lambda_2$, 
consistent with the fact that the $p$-wave $\Lambda\Lambda$ interaction 
does not contribute to the ground state of an ordinary double-$\Lambda$ system. 
The local potential $U_\Lambda(r)$, however, shows an apparent 
$\lambda_2$ dependence. 
This does not contradict the vanishing of the $p$-wave contribution 
to the $(1s)^2$ configuration, because the plotted quantity is only 
the local central part of the Skyrme mean field. 
The momentum-dependent $\Lambda\Lambda$ terms also modify the effective-mass 
of the one-body Hamiltonian, and the contributions cancel 
for a pair of $\Lambda$ hyperons occupying the same $1s$ orbit.

A qualitatively different behavior appears for $N_\Lambda=8$, 
where the $1p$ shell is occupied. 
The repulsive $\lambda_2$ term then contributes through relative 
$p$-wave components involving the $1p$ orbit. 
As $\lambda_2$ increases, the $\Lambda$ mean field becomes shallower 
and the $1p$ orbit is pushed closer to the continuum. 
Consequently, the $\Lambda$ density distribution extends to larger radii, 
which is the microscopic origin of the strong $\lambda_2$ dependence of 
$r_\Lambda$ seen in Fig.~\ref{fig:O16+mul-L}.

Next, we examine medium and heavy cores, $^{40}$Ca and $^{208}$Pb. 
Since the dependence on $\lambda_3$ within 
$0\leq\lambda_3\leq 1000$ MeV\,fm$^6$ 
is found to be minor compared with the $\lambda_2$ dependence, 
we show only the results for the LL4$(\lambda_2,0)$ parameter sets 
in the following.

\begin{figure*}
\includegraphics[width=.6\linewidth, page=1]{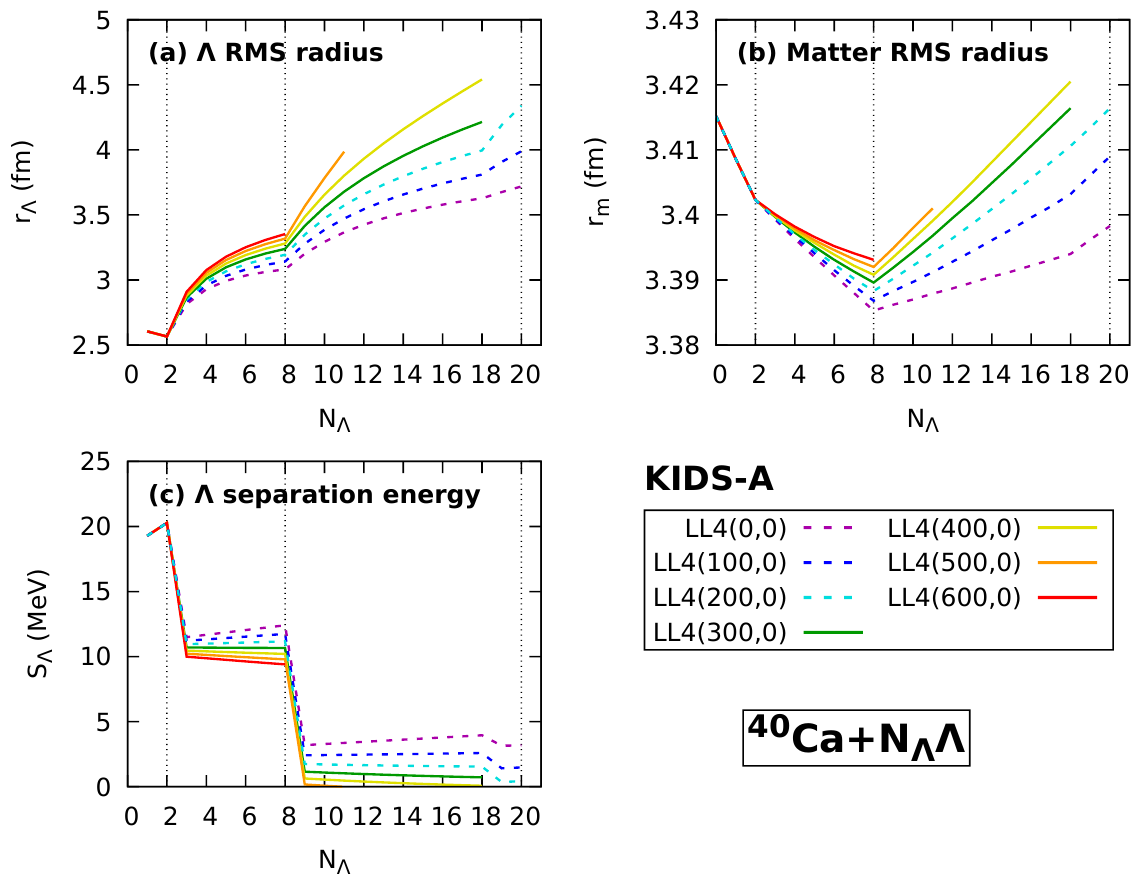}
\caption{HF results for multi-$\Lambda$ hypernuclei with the
$^{40}$Ca core, calculated with KIDS-A-Y4-LL4$(\lambda_2,0)$.
Panels show (a) the $\Lambda$ rms radius $r_\Lambda$,
(b) the matter rms radius $r_m$, and
(c) the $\Lambda$ separation energy $S_\Lambda$ as functions of
$N_\Lambda$.
The line color distinguishes the value of $\lambda_2$.
Dashed curves correspond to parameter sets disfavored by observations of
massive ($\sim2M_\odot$) neutron stars \cite{TaHyCh26}.
Vertical dotted lines indicate the closures of the $1s$, $1p$, and
$2s1d$ $\Lambda$ shells at $N_\Lambda=2$, 8, and 20, respectively.}
\label{fig:obs-Ca40+mul-L}
\end{figure*}

Figure~\ref{fig:obs-Ca40+mul-L} shows the HF results for 
multi-$\Lambda$ hypernuclei with the $^{40}$Ca core.
The overall trend is similar to that found for the $^{16}$O core. 
The matter radius changes only weakly, 
indicating that the core-polarization effect remains at the level of about 1\%. 
By contrast, the $\Lambda$ radius and the $\Lambda$ separation energy 
show an increasingly visible $\lambda_2$ dependence as $N_\Lambda$ increases, 
reflecting the cumulative role of the $\Lambda\Lambda$ interaction. 
The stepwise behavior of $S_\Lambda$ also indicates that the 
$\Lambda$ shell structure is clearly retained in the multi-$\Lambda$ system. 
Near the $\Lambda$ drip line, the highest occupied $\Lambda$ orbit 
becomes weakly bound, and the $\Lambda$ radius increases rapidly. 
For the $^{40}$Ca core, the $\Lambda$ drip is reached in the 
$sd$-shell region for the neutron-star-compatible parameter sets shown by solid curves.

\begin{figure*}
\includegraphics[width=.6\linewidth, page=1]{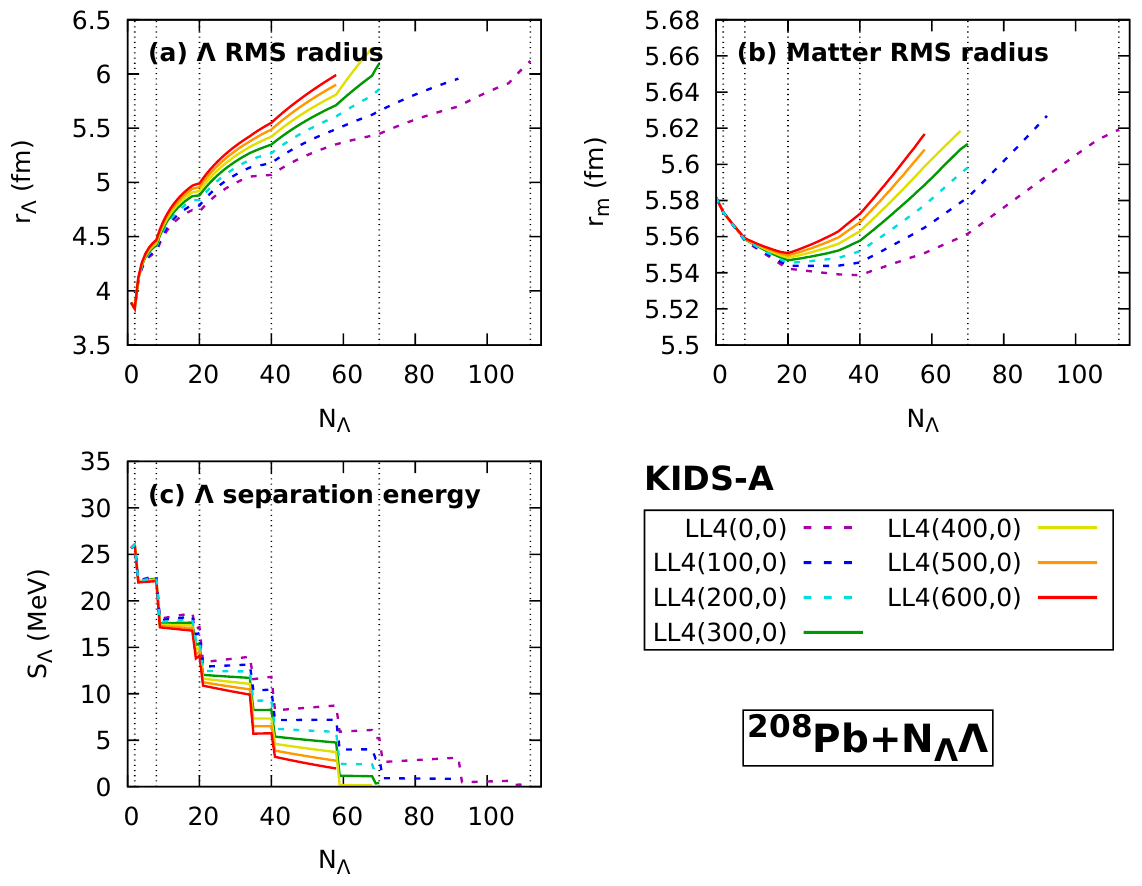}
\caption{Same as Fig.~\ref{fig:obs-Ca40+mul-L}, but for multi-$\Lambda$
hypernuclei with the $^{208}$Pb core.
Panels show (a) $r_\Lambda$, (b) $r_m$, and (c) $S_\Lambda$
as functions of $N_\Lambda$.
The vertical dotted lines indicate representative $\Lambda$ shell closures,
and dashed curves denote parameter sets disfavored by observations of
massive ($\sim2M_\odot$) neutron stars.}
\label{fig:obs-Pb208+mul-L}
\end{figure*}

Figure~\ref{fig:obs-Pb208+mul-L} shows the corresponding results 
for the heavy $^{208}$Pb core, again for $\lambda_3=0$. 
The qualitative behavior is the same as in the lighter systems: 
the matter radius is only weakly affected by the addition of 
$\Lambda$ hyperons, whereas the $\Lambda$ radius and the 
$\Lambda$ separation energy become increasingly sensitive to 
$\lambda_2$ as $N_\Lambda$ increases. 
The shell structure is reflected in the stepwise decrease of 
$S_\Lambda$, while the $\Lambda$ drip line itself exhibits a more 
pronounced $\lambda_2$ dependence than in lighter systems. 
This trend is associated with the higher density of $\Lambda$ 
single-particle levels in the heavy system, 
which makes the highest occupied orbit more susceptible to the 
$\lambda_2$-induced upward shift toward the continuum. 
As a result, once the drip-line region is approached, 
the $\Lambda$ radius increases rapidly.

We should note that in medium and heavy systems the 
$\Lambda$ drip line may appear in the middle of a major shell. 
In such cases, the last occupied $\Lambda$ orbit is weakly bound 
and the quantitative position of the drip line can be sensitive to 
pairing correlations and continuum effects. 
The present HF calculation therefore provides a reference estimate 
of the drip-line behavior in the framework without pairing, 
rather than a definitive determination of the drip line. 
A more quantitative treatment of these open-$\Lambda$-shell systems 
would require HFB calculations with $\Lambda\Lambda$ pairing, 
which are left for future work.

The corresponding calculations with KIDS-D-Y4-LL4$(\lambda_2,\lambda_3)$ lead to 
qualitatively similar results for the finite multi-$\Lambda$ nuclei. 
At fixed $(\lambda_2,\lambda_3)$, the trends of $r_\Lambda$, $r_m$, $S_\Lambda$, 
and the $\Lambda$ drip-line behavior are essentially the same as those obtained with KIDS-A. 
The main difference lies in the neutron-star-compatible parameter region: 
because the $NN$-sector equation of state of KIDS-D is stiffer, weaker $\Lambda\Lambda$ 
repulsion is sufficient to satisfy the neutron-star constraints, and the acceptable region extends 
further toward smaller $\lambda_2$ and/or $\lambda_3$.

\section{Summary and outlook}
\label{sec:summary}

In this work, we have investigated finite multi-$\Lambda$ hypernuclei 
with a Skyrme-type $\Lambda\Lambda$ interaction constrained by data on 
double-$\Lambda$ hypernuclei and neutron stars. 
The $s$-wave sector of the $\Lambda\Lambda$ interaction was fixed by 
double-$\Lambda$ hypernuclear data supplemented by pseudodata from 
core+$2\Lambda$ three-body calculations, while the $p$-wave term 
and the density-dependent term were restricted through their effects 
on neutron-star matter. 
Using this constrained framework, we have examined both homogeneous hyperonic matter 
around the normal density and finite multi-$\Lambda$ hypernuclei 
within the spherical HF approach.

The analysis of homogeneous hyperonic matter around the normal density has shown that 
the nucleonic density minimizing the energy per baryon remains close to 
the normal saturation density even when a finite $\Lambda$ density is present. 
In contrast, the $\Lambda$ chemical potential and the corresponding 
$\Lambda$ drip point depend strongly on the repulsive $p$-wave parameter 
$\lambda_2$, whereas the dependence on the density-dependent parameter 
$\lambda_3$ is relatively weak within its present range 
for the densities relevant to finite nuclei. 
This already suggests that finite multi-$\Lambda$ systems provide a favorable 
environment in which the effects of the $p$-wave $\Lambda\Lambda$ interaction 
can be isolated without inducing a large rearrangement of the nucleonic core.

This expectation is confirmed by the HF results for 
multi-$\Lambda$ hypernuclei of doubly closed stable cores 
from light to heavy nuclei. 
The matter radius changes only weakly as $\Lambda$ hyperons are added, 
indicating that the core-polarization effect remains small. 
By contrast, the $\Lambda$ radius, the $\Lambda$ separation energy, 
and the $\Lambda$ single-particle structure show a clear dependence on 
$\lambda_2$, and this dependence generally becomes stronger as $N_\Lambda$ 
increases. 
This trend reflects the cumulative contribution of the $\Lambda\Lambda$ 
interaction as more $\Lambda$ hyperons occupy the finite system.

A second and distinct effect appears near the $\Lambda$ drip line. 
When the last occupied $\Lambda$ orbit approaches the continuum, 
the repulsive $p$-wave term shifts it upward and can produce a weakly bound 
state with an extended radial distribution. 
As a result, the $\Lambda$ radius may increase rapidly near the threshold, 
which should be distinguished from the more gradual enhancement of the 
$\lambda_2$ dependence with increasing $N_\Lambda$. 
The $^{16}$O-core system provides a clear example of this mechanism, 
while similar behavior is found also in heavier stable cores once the 
highest occupied $\Lambda$ orbit becomes weakly bound.
In medium and heavy systems, the resulting HF drip-line positions
should be regarded as reference estimates, especially
when the last bound configuration lies in the middle of a 
shell where pairing and continuum effects may become important.

Within the present range of $\lambda_3$, the density-dependent term has 
a relatively modest impact on finite multi-$\Lambda$ hypernuclei at densities 
around $\rho_0$, compared with the effects induced by $\lambda_2$. 
This indicates that stable finite multi-$\Lambda$ systems are particularly 
useful for isolating the role of the $p$-wave $\Lambda\Lambda$ interaction, 
whereas the density-dependent term is expected to play a more important role 
in the higher-density regime relevant to neutron stars.

Exploratory calculations for neutron-rich systems are presented in the Appendix. 
They show qualitative features consistent with previous HFB study \cite{ZhHiSa25}, 
including the possibility that diffuse mean fields in neutron-rich cores 
facilitate the appearance of weakly bound higher-shell $\Lambda$ states. 

The present results suggest several directions for future work. 
First, it will be important to extend the present HF analysis 
to HFB calculations in order to assess possible 
$\Lambda\Lambda$ pairing effects, especially in weakly bound and neutron-rich systems. 
Second, it is a natural extension of the present work to investigate deformed multi-$\Lambda$ hypernuclei, 
especially once pairing correlations are incorporated. 
Such a study would clarify how the constrained $\Lambda\Lambda$ interaction 
affects deformation energies, shell structure, and the localization or delocalization 
of $\Lambda$ hyperons in open-shell systems \cite{Ta19,LCZR24}. 
Third, if future experimental information on multi-$\Lambda$ systems becomes available, 
observables such as masses, separation energies, shell gaps, and spatial distributions 
may provide valuable constraints on the $p$-wave sector of the $\Lambda\Lambda$ interaction. 
In this sense, finite multi-$\Lambda$ hypernuclei and neutron-star matter 
should be viewed as complementary probes of the $\Lambda\Lambda$ functional: 
the former are sensitive mainly to the $p$-wave term around the normal density, 
while the latter provide an essential filter on the higher-density behavior, 
including the density-dependent repulsion.

\section*{Acknowledgments}
{
Y. T. acknowledges support from the Basic Science Research Program of the National Research Foundation of Korea (NRF) under grants No.~RS-2024-00361003, RS-2024-00460031, and RS-2021-NR060129.
Work of C. H. H. was supported by the NRF research grant No. 2023R1A2C1003177.
Work of M. K. C. was supported by the NRF research grant No. RS-2025-16071941
and No. RS-2024-00460031.
}

\appendix

\section{Neutron-rich cores}
\label{app:n-rich}

To compare with the HFB results of Ref. \cite{ZhHiSa25}, 
we show in this Appendix our HF results for neutron-rich $^{24}$O and $^{66}$Ca cores.

\begin{figure}
\includegraphics[width=\linewidth, page=1]{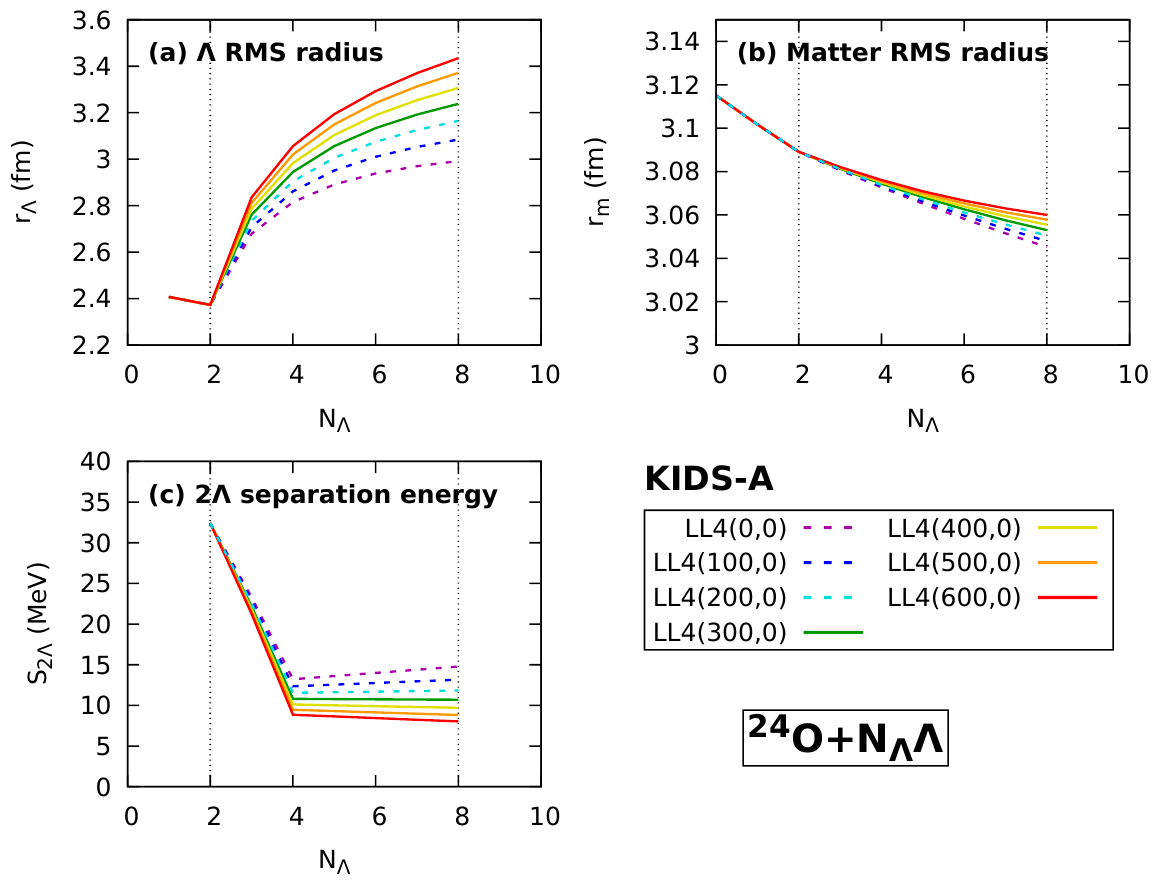}
\caption{HF results for multi-$\Lambda$ hypernuclei with the
neutron-rich $^{24}$O core, calculated with KIDS-A-Y4-LL4$(\lambda_2,0)$.
Panels show (a) the $\Lambda$ rms radius $r_\Lambda$, (b) the matter rms
radius $r_m$, and (c) the two-$\Lambda$ separation energy $S_{2\Lambda}$
as functions of $N_\Lambda$.
The line color distinguishes the value of $\lambda_2$, and dashed curves
correspond to parameter sets disfavored by observations of massive
($\sim2M_\odot$) neutron stars.
Vertical dotted lines indicate the $\Lambda$ shell closures at $N_\Lambda=2$ and 8.}
\label{fig:obs-O24+mul-L}
\end{figure}

\begin{figure}
\includegraphics[width=\linewidth, page=1]{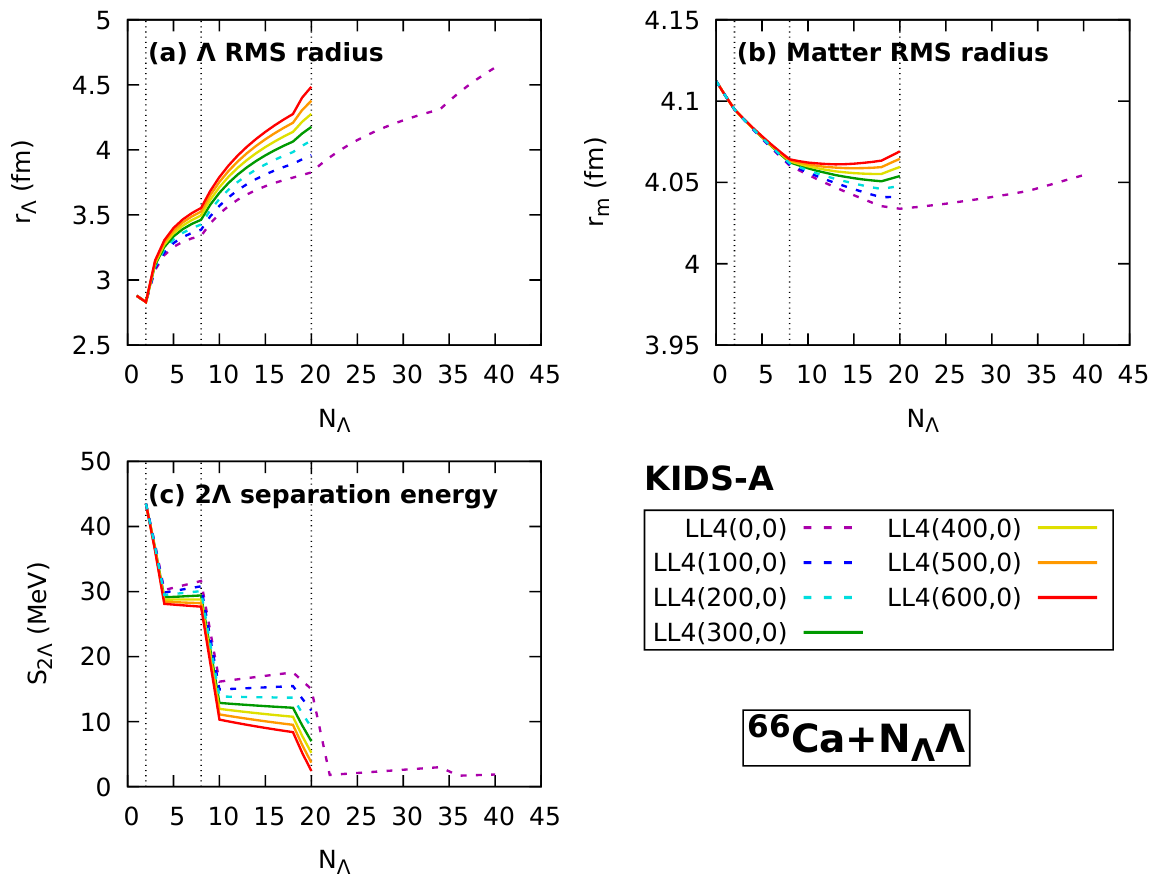}
\caption{Same as Fig.~\ref{fig:obs-O24+mul-L}, but for multi-$\Lambda$
hypernuclei with the neutron-rich $^{66}$Ca core.
The vertical dotted lines indicate the $\Lambda$ shell closures at
$N_\Lambda=2$, 8, and 20.}
\label{fig:obs-Ca66+mul-L}
\end{figure}

Figures~\ref{fig:obs-O24+mul-L} and \ref{fig:obs-Ca66+mul-L} show
the corresponding $\Lambda$ radii, matter radii, and two-$\Lambda$ separation energies.
The qualitative behavior is consistent with the HFB study of Zhang
\textit{et al.}~\cite{ZhHiSa25}: 
the drip-line behavior is sensitive to the $p$-wave $\Lambda\Lambda$ interaction, 
especially when weakly bound higher-shell $\Lambda$ orbits appear in a diffuse 
neutron-rich mean field. 
Within the neutron-star-compatible range of the present KIDS-A-Y4-LL4 parameter sets, 
the last bound configurations are still $N_\Lambda=8$ for the $^{24}$O core 
and $N_\Lambda=20$ for the $^{66}$Ca core. 
Larger extensions of the drip line occur only for parameter sets that are
outside the range favored by the neutron-star constraints
used in this work.
These results should be regarded as qualitative, because pairing and
continuum effects, which are not included in the present HF calculations, can
be important for weakly bound open-shell $\Lambda$ configurations.

\end{document}